\documentclass{jpp}
\usepackage{graphicx}
\graphicspath{{Figures/}}
\usepackage{caption}
\usepackage{subcaption}
\usepackage{float}
\usepackage{hyperref}
\usepackage[utf8]{inputenc}
\usepackage[T1]{fontenc}
\usepackage{amsmath}

\shorttitle{PIC code comparison}
\shortauthor{Banerjee et al.}

\title{Particle-In-Cell Code Comparison for Ion Acceleration: EPOCH and Smilei}

\author{Soham Banerjee\aff{1,2}
  \corresp{\email{sbaner18@ur.rochester.edu}},
  Joseph R. Smith\aff{3}, 
 Chris Orban\aff{4} 
  }

\affiliation{\aff{1}Department of Physics, Birla
Institute of Technology and Science, Pilani-333031, Rajasthan, India.
\aff{2}Department of Physics and Astronomy, University of Rochester, Rochester, NY 14627, USA.
\aff{3}Department of Physics, Marietta College, Marietta, OH, 45750, USA
\aff{4} Department of Physics, Ohio State University, 191 W Woodruff Ave, Columbus, OH 43210}
\begin{document}

\maketitle

\begin{abstract}
Particle-in-Cell (PIC) codes are a popular tool to model laser-plasma interactions. Many different PIC codes already exist, and many new PIC codes are being developed constantly. It is therefore important to compare different PIC codes to ascertain which code is best suited for a particular kind of physical problem. In a paper by \cite{smith2021} they compared three different codes on a problem relating to proton acceleration in the Target Normal Sheath Acceleration regime from a normal incidence ultra-intense laser pulse. \cite{smith2021} included in their study the widely used EPOCH code. However, they did not include results from the Smilei code, which is another popular PIC code in the plasma community with a variety of features and physics packages. In the present work, we compare the Smilei code to the EPOCH code for the same test case as \cite{smith2021}. Broadly we find the two codes to be highly consistent with agreement in total ion, electron, and field energy at a percent level or better. The electron and ion energy distribution functions agree well at lower energies and the differences at higher energies (e.g. because of the finite number of macroparticles) are similar to what \cite{smith2021} saw for other codes. We found that Smilei consumed 25\% more RAM than EPOCH did but the execution time was 30\% less for Smilei on one processor. We include the input files to encourage future comparisons.
\end{abstract}

\section{Introduction}
Numerical modeling is a crucial tool for studying various phenomena in plasma physics. Due to non-linear interactions between a vast number of physical processes in plasmas, there is no complete analytical theory that can account for all these different processes. Therefore, computer codes are widely used to model plasma behavior in various regimes, such as low temperature plasmas, magnetically confined plasmas, astrophysical plasmas, and intense laser-plasma interactions. 
In situations like this, where reliance upon numerical modeling is high and there are relatively few analytic solutions to benchmark the codes, it is important to compare the predictions of many different codes on problems of interest, and to continue these comparisons as the codes mature and new codes are developed. In this work we consider a problem related to ultra-intense laser acceleration of protons and electrons and we compare the results of two different codes that implement the Particle-in-Cell (PIC) method \citep{Dawson1962, Harlow1962, Hockney2021, Birdsall2018}. 

PIC codes have been used extensively to study Laser-Plasma interactions (LPI). Over the years there have been numerous efforts to develop PIC codes that can model a wide range phenomena. As a result, there currently exist a large number of PIC codes for LPI. An incomplete list includes EPOCH \citep{Arber2015}, Smilei\citep{Derouillat2018}, Warp-X\citep{Vay2018}, LSP\citep{Welch2004,Welch2016}, VORPAL\citep{Nieter2004}, OSIRIS\citep{Fonseca2002}, VLPL\citep{pukhov1999}, PIConGPU\citep{Burau2010}, QUICKPIC\citep{Huang2006}, AlaDyn\citep{Benedetti2008}, Chicago\citep{Thoma2017}, MAGIC\citep{Goplen1995}, PICCANTE\citep{Andrea2015}, and OOPIC\citep{VERBONCOEUR1995}. 

Although many PIC codes exist, there are relatively few studies comparing these codes. 

In the literature there have been some published code-to-code comparisons. For example, \cite{Sun2016} compared their PIC/MCC simulations with the established simulation benchmarks\citep{Turner2013} for low temperature CCRF discharges. In another study,  \cite{paul2009} compared LWFA simulation results from VORPAL, OSIRIS, and QuickPIC. \cite{dollar2013} modeled their ion acceleration experiments using OSIRIS and an in-house developed implicit code for comparison. \cite{cochran2018new} simulated ion acceleration using implicit and explicit schemes in LSP. \cite{smith2021} compared the results and performance of three commonly used PIC codes, namely EPOCH, WarpX, and LSP, and found good agreement between the results. \cite{smith2021} provided the input files for these simulations in an effort to encourage others to continue the work and add more codes and extend the comparison.  \citet{mouziouras_compare_thesis} performed an informative comparison of Smilei and EPOCH in a master's thesis, although quantitative comparisons are difficult as input files are not included and results are graphed separately with different plotting options. 

In this paper, we extend the results of \cite{smith2021} to include the Smilei code which we directly compare to predictions from the EPOCH code using the same input file from \cite{smith2021}. Like EPOCH, Smilei is also a widely used open-source PIC code used for plasma simulations. Smilei\citep{Derouillat2018} is more recent as compared to EPOCH\citep{Arber2015}, but it has gained popularity among users for all kinds of plasma simulations, especially LPI. While both EPOCH and Smilei are similar in their explicit-scheme PIC algorithms, there are a few important features unique to each code that can bring about differences in simulations. For example, Smilei has focused a lot on speeding up the PIC loop. It uses a sophisticated OpenMP-MPI hybrid parallelization scheme, along with a vectorization scheme\citep{BECK2019}, that can reduce simulation times. On the other hand, EPOCH has procedures to eliminate noise in the results, such as the $\delta f$ capability, that can help reduce noise and yield better results for plasma problems. There are also many implementation differences in both codes, which do not affect the physics, but may be attractive to certain users. For example, EPOCH uses SI units, while Smilei uses its own code units in its input files and during post-processing. The differences mentioned above, along with the popularity of both codes in the community, make it worthwhile comparing the performance of both codes. 
\\
As in \cite{smith2021}, we carry out 2D3v simulations of a case of Target Normal Sheath Acceleration (TNSA)\citep{Wilks_2001,Passoni_2010}. 
Although we do not compare to other PIC codes, the interested reader can directly compare our results to the other codes (WarpX and LSP) presented in \cite{smith2021}. We chose to focus on Smilei and EPOCH because EPOCH is well studied in \cite{smith2021} and well described in \cite{Arber2015} and it is one of the most frequently used PIC codes in the field. The Smilei code is a newer code but it is becoming more popular (e.g. \cite{meinhold_kumar_2021, Yao_etal2021,Bonvalet_etal2021,Singh_etal2022}). We performed this study without significant input or advice from either the Smilei or EPOCH development teams.

\section{Simulation setup}
\begin{figure}
         \centering
         \includegraphics[width=0.7\textwidth]{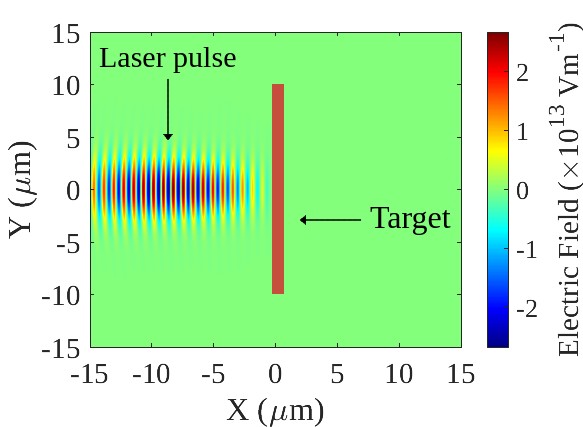}
         \caption{2D Simulation setup of the ion acceleration problem.}
         \label{fig:1}
\end{figure}

In this study we use EPOCH version 4.19.2 and Smilei version 4.7 for our simulations. The simulation setup, shown in \hyperref[fig:1]{Figure 1}, and parameters in this study closely follow the setup in \citet{smith2021}. To briefly summarize that setup, the problem uses 2D3v geometry and the simulation domain consists of a square in the XY plane centered at the origin, with the edges of the simulation each being $30~\mu$m in size. The target is a hydrogen plasma slab with a width of $5~\mu$m along the x-direction and a length of $20~\mu$m along the y-direction. The density of the target is chosen to be $n_d = 5 \cdot n_{\rm crit}$, where $n_{\rm crit} = \epsilon_0 m_e \omega^{2}/e^2$ is the
nonrelativistic critical density of the plasma, with $\epsilon_0$ being the permittivity of free space, $m_e$ being the electron mass, $\omega$ being the angular frequency of the laser, and $e$ being the electron charge. For a laser wavelength of $\lambda = 800$~nm, the density of the target comes out to be $n_d = 8.5\times10^{21}$~cm$^{-3}$. This density is sufficiently over-dense for simulating a case of TNSA, and low enough to comfortably carry out 2D-PIC simulations without requiring vast computational resources. The laser is linearly polarized in the $z$-direction(out of the plane of the simulation) and moves in the $+x$ direction starting from the boundary of the simulation domain. The laser is a Gaussian beam focused on the target (at the origin) with peak intensity of $10^{20}$~W~cm$^{-2}$ and a Full Width at Half Maximum(FWHM) of $3~\mu$m. This corresponds to a normalized potential of $a_0 = \frac{e E_0 }{m_e \omega c} = 6.8$ and a beam waist of approximately $w_0 = 2.55~\mu$m. The time profile of the laser pulse is a sinusoidal wave with a FWHM of 30 fs.\\

Likewise the numerical parameters are chosen to be the same as was assumed in \citet{smith2021}. The cell size is $20$~nm, which gives $1500\times1500$ cells in the simulation domain. We have used 100 macroparticles in each cell.

The time step is chosen as $0.04$~fs for a total simulation time of $300$~fs, which works out to $7500$ time steps. This time step satisfies the CFL condition by a factor of approximately $0.848$. The cell size also resolves the skin depth $l_{sd} = c/\omega_p = 58$~nm, where the plasma frequency $\omega_p = \sqrt{\frac{n_e e^{2}}{m_e \epsilon_0}} = 5.2\times10^{16}$~Hz. The skin depth is clearly smaller than the laser wavelength (800nm) we used, which means the skin depth is resolved with approximately 3 cells per skin depth.\\

The particles in the target are initialized using a Maxwell-Boltzmann distribution in EPOCH and a Maxwell-Juttner distribution in Smilei. Smilei uses the Maxwell-Juttner distribution as a default, which naturally reduces to the Maxwell-Boltzmann for low temperatures. Since our initial temperature $k_B T = 10~$keV is much smaller than the electron rest mass energy (0.5~MeV), there is no relativistic correction required, and the Maxwell-Juttner distribution is equivalent to the Maxwell-Boltzmann distribution. As for particle shapes, EPOCH and Smilei provide a few different options for particle shape functions. We have used second order particle shape for the macroparticles our simulations. This means that the macroparticle is taken as triangle shaped in each dimension with a width of 2 cells. The field interpolation technique is chosen as the default momentum-conserving scheme in both codes. Lastly, we have used the Yee grid option for solving Maxwell's equations and the Boris particle pusher, which are the defaults in both EPOCH and Smilei simulations. For easy reference, all these parameters can be found in the input files we have used, which we have included in the supplementary material for this paper.\\ 

\section{Results}

\subsection{Comparing Predicted Quantities}

\begin{figure}
    \centering
    \includegraphics[width=5.5cm]{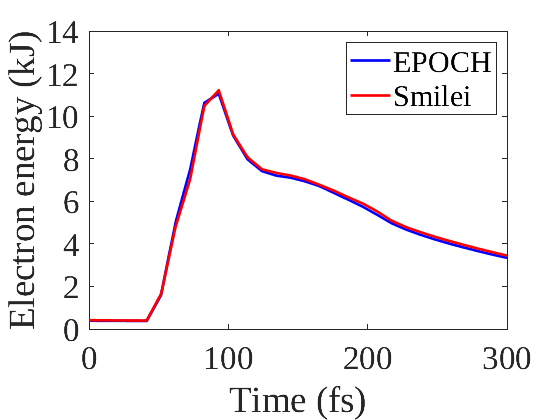}\includegraphics[width=5.5cm]{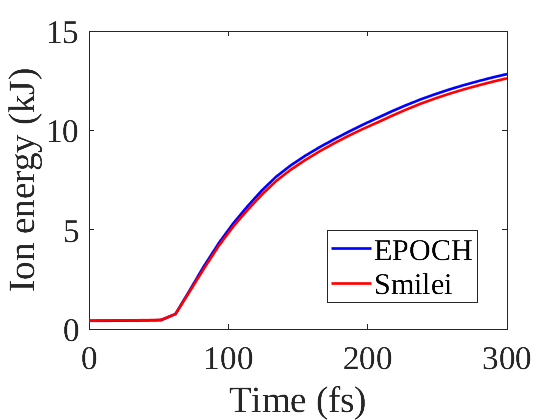}
 \includegraphics[width=5.5cm]{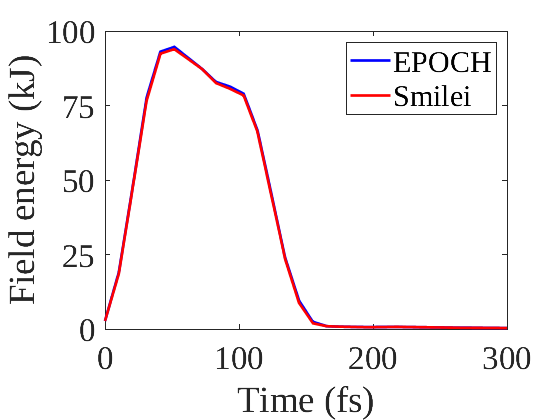}
    \caption{Comparing the time evolution of electron energy (upper left panel), ion energy (upper right panel), and field energy (lower panel) for EPOCH (blue) and Smilei (red).}
    \label{fig:energy_vs_time}
\end{figure}

In this section, we directly compare the simulation results of Smilei and EPOCH. Fig.~\ref{fig:energy_vs_time}, shows the evolution of total electron energy, ion energy, and field energy across time. The laser hits the target at about $50$~fs, at which point the electron and ion energy begin to increase. Results from both codes show strong agreement with respect to all three quantities. Comparing the evolution of electron energy over time shows very little difference -- about 0.05 kJ -- between Smilei and EPOCH towards the end of the simulation which is sub-percent level agreement. Similarly, the results for the evolution of total ion energy remain highly consistent with time, differing only by 0.2~kJ or 1.4\%  at the end of the simulation with EPOCH predicting slightly more total ion energy. 

In case of energy in the fields, both Smilei and EPOCH show very similar results with better than percent level agreement throughout the simulation.\\

\begin{figure}
     \centering
\includegraphics[width=5.5cm]{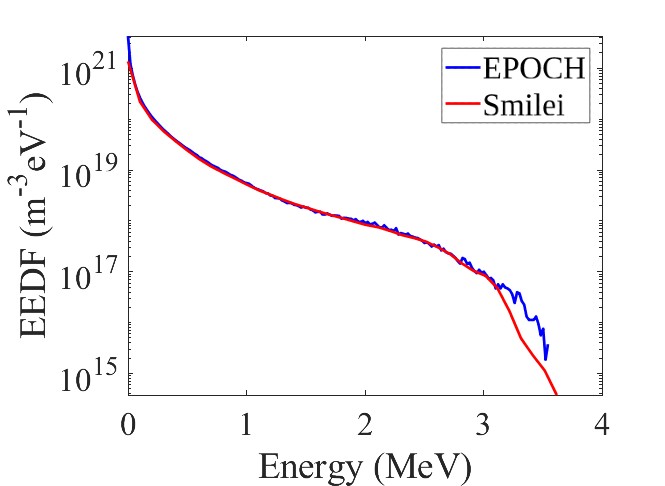}\includegraphics[width=5.9cm]{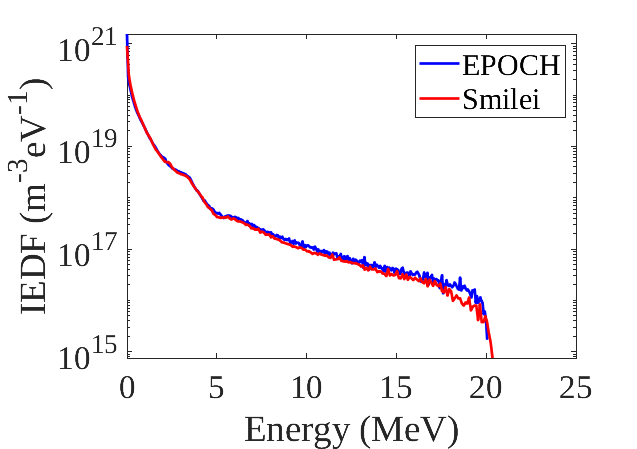}
          \caption{Comparison of electron (left panel) and ion energy (right panel) distribution functions at the end of the simulation for EPOCH (blue) and Smilei(red).}
        \label{fig:spectra}
\end{figure}

In Fig.~\ref{fig:spectra} we show the electron and ion energy distribution functions (the EEDF and IEDF respectively) at the end of the simulation (300~fs). Both codes show strong agreement in the EEDF as well as the IEDF, with slight differences towards the high energy tail of the distributions. The ion energy cut off is observed to be $20.03$~MeV in Epoch, and $20.5$~MeV in Smilei. Because of the finite number of particles per species per cell we do not expect perfect agreement at the highest energy bin. This difference of less than $0.5$~MeV in the ion energy cutoff is similar to what \cite{smith2021} found in comparing the WarpX and LSP codes to EPOCH. In our results, the EPOCH predictions for EEDF and IEDF are slightly above that of Smilei, which may be related to results in Fig.~\ref{fig:energy_vs_time} which showed that at the end of the simulation the total ion energy was 1.4\% higher in EPOCH while the total electron energy agreement was sub percent level but with EPOCH predicting slightly more total electron energy.

\begin{figure}
         \centering
         \includegraphics[width=0.65\textwidth]{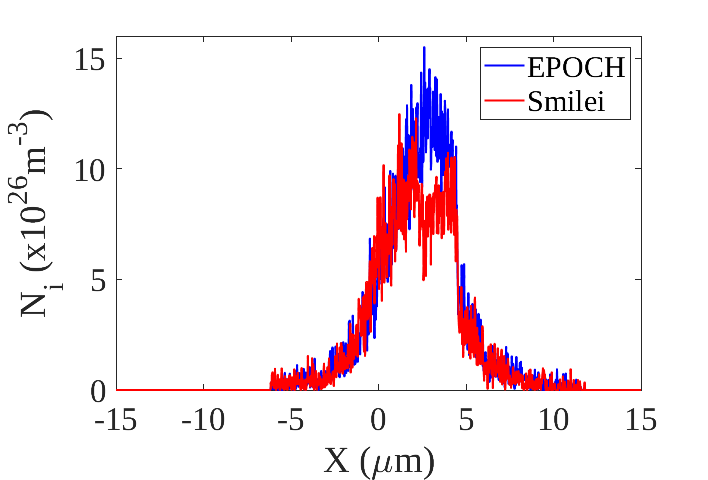}
         \caption{Comparison of ion density profiles at the end of the simulation for EPOCH(blue) and Smilei(red).}
         \label{fig:ion_density}
\end{figure}

Fig.~\ref{fig:ion_density} shows the plot of ion density at the end of the simulation across the center of the simulation domain, i.e. along the $y=0$ line. There is a difference in the ion density profiles of both codes, with Smilei showing a dip in ion density at the center of the expanded target. These results mirror what \cite{smith2021} found in their Figure 5 which concluded that the ion density profiles do differ at this level between codes.

\subsection{Performance and Memory Consumption}

\begin{figure}
         \centering
         \includegraphics[width=0.65\textwidth]{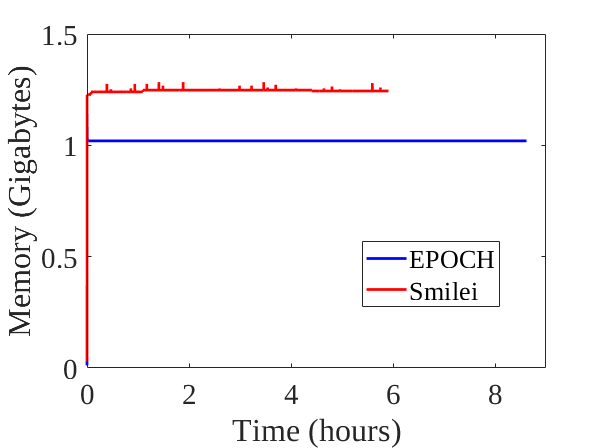}
         \caption{Comparison of the memory consumption by EPOCH(blue) and Smilei (red and green) simulations.}
         \label{fig:memory}
\end{figure}

Fig.~\ref{fig:memory} shows the memory consumed by both codes against the time. We did not include any diagnostics in these runs, which would increase memory usage of both codes by a small amount.  Smilei uses a hybrid OpenMP-MPI parallel scheme by default, however our tests were all single core so we were not well positioned to see if this hybrid scheme is helpful. Therefore, for our runs, we compiled Smilei without OpenMP. EPOCH uses an MPI parallel scheme without OpenMP acceleration.\\

Compared to EPOCH, Smilei used more memory throughout the run, around 1.25~GB, whereas EPOCH used around 1~GB. The Smilei simulation took about 6 hours to complete, while the Epoch simulation ran for 8 hours and 37 minutes. The Smilei simulation therefore used 25\% more memory than EPOCH but ran about 30\% faster. This result did not involve any other efforts to improve the performance besides removing the OpenMP compiler flag.

\section{Discussion}

We find a high degree of agreement between the EPOCH and Smilei codes. Observed differences are similar to what \cite{smith2021} found in comparing the WarpX and LSP codes to the EPOCH code. Differences in the electron and ion energy distribution functions (which are perhaps the most important quantities from an experimental point of view) are predominantly in the high energy tail where the finite number of macroparticles in the simulation will naturally cause fluctuations in the measurement. Although we did not perform our own tests, with this in mind \cite{smith2021} performed a number of EPOCH simulations to measure the range of results for the ion density distribution function as highlighted in their Figure 3. Our ion energy cutoffs of 20~MeV and 20.5~MeV fall within the range that was observed there. 

In terms of performance and memory consumption, we found that Smilei (with OpenMP disabled) used 25\% more memory than EPOCH but ran 30\% faster. Tests that included the OpenMP compiler flag with Smilei had similar performance but 65\% more RAM consumption than EPOCH. However, this extra RAM usage by OpenMP is irrelevant to our purpose since we have only used 1 core. 
Given that the codes produce similar results, this does suggest that users who have access to machines with more memory may want to use Smilei to take advantage of the extra speed. However, we caution that our results were obtained with a single core, so the performance results may not straightforwardly transfer to a parallel cluster.  

It is worth mentioning again that Smilei uses a hybrid OpenMP-MPI parallelization scheme. This means that in addition to traditional MPI parallelization, Smilei can use OpenMP to smartly allocate resources inside each MPI process, which in turn reduces simulation time for a constant number of cores (as described in section 4.3 of \citep{Derouillat2018}). Due to this additional capability in Smilei, one may expect Smilei to perform faster in parallel tests compared to EPOCH as the number of cores increases. Although we have not run the codes using multiple cores, we encourage interested users to perform their own tests. Interestingly, \citet{mouziouras_compare_thesis} did find Smilei to run faster than EPOCH when simulating laser wakefield acceleration on tens of CPUS in their tests.  

However, this is not to say that one code is better than the other. Both EPOCH and Smilei have capabilities and packages that can be suited to particular plasma physics problems. For example, EPOCH has a $\delta f$ capability, which uses a background distribution function to significantly reduce numerical noise in simulation results. This capability is helpful when simulating turbulence, such as in tokamaks. EPOCH has also been more frequently used to model QED effects in plasmas\citep{Duff_2019,Han_2022,MacLeod2023}, even though both EPOCH and Smilei have QED physics packages. On the other hand, Smilei has a variety of options to reduce simulation times. In addition to the hybrid parallelization scheme, it has also recently added a vectorization capability\citep{BECK2019} that can further speed up simulations depending upon the nature of the problem and the number of macroparticles.

We performed our tests without significant input from the Smilei or EPOCH development teams and the EPOCH simulation setup was identical to that used in \cite{smith2021}. This was intentional, but it means that there could be ways to improve the speed or memory consumption of the codes that we were not aware of. Interested users should contact the development teams for these codes for the latest advice.

\section{Conclusion}

We performed Particle-in-Cell simulations of proton acceleration via the TNSA mechanism from an ultra-intense laser pulse irradiating a slab at normal incidence using the Smilei and EPOCH codes. Comparing single core performance we found that Smilei (with OpenMP disabled) consumed 25\% more memory than EPOCH, but outperformed EPOCH in terms of execution time by 30\%. The proton and electron energy distributions were very similar and with a similar high energy cutoff. Overall, the agreement between Smilei and EPOCH is similar to the agreement between EPOCH and the WarpX and LSP codes described in \citet{smith2021}.

\section*{Acknowledgements}

 We would like to thank Frederic Perez from the Smilei team for useful feedback. We also thank the Smilei team for keeping their chatroom active, which was helpful during code installation and other technical questions. The Smilei code is governed by the CeCILL-B license under French law and abiding by the rules of distribution of free software. You can use, modify, and/or redistribute the software under the terms of the CeCILL-B license as circulated by CEA, CNRS and INRIA at the following URL \url{http://www.cecill.info}. The code EPOCH used in this work was in part funded by the UK EPSRC grants EP/G054950/1, EP/G056803/1, EP/G055165/1 and EP/ M022463/1.

\bibliographystyle{jpp}

\bibliography{jpp-instructions}

\end{document}